\documentclass{nature}
\newcommand {\gtrsim} {\ {\raise-.5ex\hbox{$\buildrel>\over\sim$}}\ }
\newcommand {\ltrsim} {\ {\raise-.5ex\hbox{$\buildrel<\over\sim$}}\ }
\usepackage{psfig}
\usepackage{graphicx}
\usepackage{deluxetable}
\usepackage{amsmath}
\usepackage{graphicx}
\usepackage{caption}
\usepackage{subcaption}

\title{Binary orbits as the driver of $\gamma$-ray emission and mass ejection in classical novae}

\author{Laura Chomiuk$^{1}$, Justin D.~Linford$^{1}$, Jun Yang$^{2,3,4}$, T.~J.~O'Brien$^{5}$, Zsolt Paragi$^{3}$, Amy J.~Mioduszewski$^{6}$, R.~J.~Beswick$^{5}$, C.~C.~Cheung$^{7}$,
Koji Mukai$^{8,9}$, Thomas Nelson$^{10}$, Val\'erio A.~R.~M.~Ribeiro$^{11}$, Michael P.~Rupen$^{12,6}$, J.~L.~Sokoloski$^{13}$, Jennifer Weston$^{13}$, Yong Zheng$^{13}$, 
Michael~F.~Bode$^{14}$, Stewart Eyres$^{15}$, 
Nirupam Roy$^{16}$, Gregory B.~Taylor$^{17}$}

\begin{document}
\spacing{1}
\maketitle
\begin{affiliations}
\item Department of Physics and Astronomy, Michigan State University, East Lansing, MI 48824, USA
\item Department of Earth and Space Sciences, Chalmers University of Technology, Onsala Space Observatory, SE-439 92 Onsala, Sweden
\item Joint Institute for VLBI in Europe, Postbus 2, NL-7990 AA Dwingeloo, the Netherlands
\item Shanghai Astronomical Observatory, Chinese Academy of Sciences, 80 Nandan Road, 200030 Shanghai, China
\item Jodrell Bank Centre for Astrophysics, Alan Turing Building, University of Manchester, Manchester, M13 9PL, UK
\item National Radio Astronomy Observatory, P.O. Box O, Socorro, NM 87801, USA
\item Space Science Division, Naval Research Laboratory, Washington, DC 20375-5352, USA
\item Department of Physics, University of Maryland, Baltimore County, 1000 Hilltop Circle, Baltimore MD 21250, USA
\item CRESST and X-ray Astrophysics Laboratory, NASA/GSFC, Greenbelt MD 20771 USA
\item School of Physics and Astronomy, University of Minnesota, 115 Church St SE, Minneapolis, MN 55455, USA
\item Astrophysics, Cosmology and Gravity Centre, Department of Astronomy, University of Cape Town, Private Bag X3, Rondebosch 7701, South Africa
\item National Research Council, Herzberg Astronomy and Astrophysics, 717 White Lake Road, P.O. Box 248, Penticton, BC  V2A 6J9, Canada 
\item Columbia Astrophysics Laboratory, Columbia University, New York, NY, USA
\item Astrophysics Research Institute, Liverpool John Moores University, IC2, Liverpool Science Park, 146 Brownlow Hill, Liverpool L3 5RF, UK
\item Jeremiah Horrocks Institute for Mathematics, Physics, \& Astronomy, University of Central Lancashire, Preston PR1 2HE, UK
\item Max-Planck-Institut f\"{u}r Radioastronomie, Auf dem H\"{u}gel 69, D-53121 Bonn, Germany
\item Department of Physics and Astronomy, University of New Mexico, MSC07 4220, Albuquerque, NM 87131-0001, USA
\end{affiliations}

\begin{abstract}

Classical novae are the most common astrophysical thermonuclear explosions, occurring on the surfaces of white dwarf stars accreting gas from companions in binary star systems\cite{Gehrz98}.
 Novae typically expel $\sim$10$^{-4}$ solar masses of material at velocities exceeding 1,000 kilometres per second. However, the mechanism of mass ejection in novae is poorly understood, and could be dominated by the impulsive flash of thermonuclear energy\cite{Starrfield72}, prolonged optically thick winds\cite{Kato94}, or binary interaction with the nova envelope\cite{MacDonald80}. Classical novae are now routinely detected in gigaelectronvolt $\gamma$-ray wavelengths\cite{Cheung13}, suggesting that relativistic particles are accelerated by strong shocks in the ejecta. Here we report high-resolution radio imaging of the $\gamma$-ray-emitting nova V959~Mon. We find that its ejecta were shaped by the motion of the binary system: some gas was expelled rapidly along the poles as a wind from the white dwarf, while denser material drifted out along the equatorial plane, propelled by orbital motion\cite{Soker89,Porter98}. At the interface between the equatorial and polar regions, we observe synchrotron emission indicative of shocks and relativistic particle acceleration, thereby pinpointing the location of $\gamma$-ray production. Binary shaping of the nova ejecta and associated internal shocks are expected to be widespread among novae\cite{Shankar91}, explaining why many novae are $\gamma$-ray emitters\cite{Cheung13}.
\end{abstract}

\noindent The identification of the $\gamma$-ray transient J0639+0548, detected by NASA?s Fermi Gamma-ray Space Telescope, with the classical nova V959 Mon\cite{Cheung13} was a surprise, because gigaelectronvolt $\gamma$-rays are produced via the Inverse Compton, the pion production mechanism, or both\cite{Dubus13}, requiring a population of relativistic particles which had not been predicted or observed in normal classical novae. Gigaelectronvolt $\gamma$-rays had only been reported from one nova previous to V959~Mon, in a system with an unusual Mira giant companion, dense circumbinary material, and thereby strong shock interaction between the nova ejecta and surroundings\cite{Abdo10}. The white dwarf in V959~Mon, on the other hand, has a main sequence companion and therefore a low-density circumbinary environment\cite{Munari13, Hoard14}, and so there is no apparent mechanism for diffusive shock acceleration in an interaction with surrounding material.

\noindent The $\gamma$-ray emission from V959 Mon was discovered on 2012 June 19 (day 0) and lasted $\sim$12 d, showing a soft-spectrum continuum\cite{Cheung13}. Little is known about V959~Mon during the period of $\gamma$-ray emission owing to its solar conjunction in the first few months of outburst, which prevented optical observations; the transient was not even identified as a nova until 56 d after $\gamma$-ray discovery\cite{Cheung13}. However, we obtained early radio observations coincident with the Fermi detections using the Karl G.~Jansky Very Large Array (VLA), just 12 and 16 d after discovery (Fig.\ 1). These observations span a frequency range 1--6 GHz and show a flat radio spectrum ($\alpha \approx -0.1$, where $S_{\nu} \propto \nu^{\alpha}$; $\nu$ is the observing frequency and $S_{\nu}$ is the flux density at this frequency). This spectral index is much more consistent with synchrotron radiation than the expected optically-thick emission from warm nova ejecta\cite{Seaquist08, Roy13} ($\alpha \approx 2$ is predicted and observed in V959~Mon at later times; Fig.\ 1; Extended Data Fig.\ 1). 

\noindent Like gigaelectronvolt $\gamma$-rays, synchrotron emission requires a population of relativistic particles, so we can use this radio emission as a tracer of $\gamma$-ray production that lasts longer and enables much higher spatial resolution than do the $\gamma$-rays themselves. The location of the $\gamma$-ray producing shocks was revealed by milliarcsecond-resolution radio imaging using very-long-baseline interferometric (VLBI) techniques, which are sensitive to high-surface-brightness synchrotron emission. VLBI observations were achieved with the European VLBI Network (EVN) and Very Long Baseline Array (VLBA), and spanned 2012 September 18 to October 30 (91--133 d after $\gamma$-ray discovery; 2--7 mas resolution; 1 mas $\approx 2 \times 10^{13}$ cm at the distance of V959~Mon\cite{Munari13}; Extended Data Table 3). The first VLBI epoch revealed two distinct knots of emission separated by 36 mas, which we subsequently observed to travel away from one another at an estimated rate, $\sim$0.4 mas d$^{-1}$ (Fig.\ 2a). In addition, a third radio component appeared in our imaging from day 113. The brightest component was slightly resolved by the VLBA on day 106 (Extended Data Fig.\ 2), and had a peak brightness temperature, $\sim2 \times 10^6$ K, indicative of non-thermal emission (X-ray observations from around this time\cite{Nelson12} imply that hot shocked thermal gas can only account for $<$10\% of the radio flux density seen in the VLBI knots; see Supplementary Information). 

\noindent Observations made with the e-MERLIN array (54 mas resolution) just before the first VLBI epoch shows that the compact VLBI components were embedded in a larger-scale structure which was mostly extended east-west (and not detected in the VLBI imaging, because these high-resolution arrays with widely separated antennas are not sensitive to such diffuse emission; Fig.\ 2b). This diffuse emission is interpreted as thermal bremsstrahlung from the bulk of the nova ejecta\cite{Seaquist08, Roy13}. Whereas the 5 GHz flux density detected in our VLBI imaging was roughly constant or declining with time, the flux density detected in the lower-resolution observations rapidly increased during this period (Fig.\ 1). The VLBI knots comprised 19\% and 9\% of the total 5 GHz flux density on days 91 and 117 respectively, implying that over time, the synchrotron emission becomes overwhelmed by thermal emission from the warm ejecta. Although synchrotron radio emission has been detected from outbursting novae with red giant companions and dense circumbinary material\cite{Seaquist89, O'Brien06, Kantharia14}, it has not previously been securely identified in novae with main sequence companions\cite{Seaquist08}, owing to a paucity of high-resolution radio imaging enabling components of differing surface brightness to be clearly distinguished.

\noindent The expanding thermal ejecta were resolved with the VLA when it entered its high-resolution A configuration. An image from 2012 October 23 (day 126; 43 mas resolution) shows that the ejecta have expanded and assumed a clearly bipolar geometry consistent with analyses of optical spectral line profiles\cite{Ribeiro13, Shore13} (Fig.\ 2c). The apparent geometry of the ejecta is conveniently simplified, because we view the orbital plane of V959~Mon edge-on\cite{Page13}.
Our imaging illustrates that the VLBI knots were not simple jet-like protrusions from the thermal ejecta. First, there were three VLBI knots when only two are expected from a bipolar jet structure. Second, the major axis of the thermal ejecta (directed east--west) was not well aligned with the expansion of the VLBI knots, but offset by 45$^{\circ}$. In addition, the thermal ejecta expanded faster than the VLBI knots (0.64 mas/day in diameter; Extended Data Fig.\ 3). Finally, because the warm thermal ejecta were optically thick at the time of this imaging, the VLBI knots were superimposed around the edges of the ejecta, appearing to surround the two thermal lobes.
 
\noindent The origin of the compact radio knots was clarified when we revisited V959~Mon sixteen months later, when the VLA was next in A configuration (2014 February 24; day 615; Fig.\ 2d). The much-expanded thermal ejecta maintained a bi-lobed morphology---but the axis of elongation had rotated so that the brightest regions were oriented north--south, perpendicular to the outflow observed in 2012. The position angle of the VLBI knots lay roughly halfway between that of the early and late axes of ejecta expansion (Fig.\ 2).

\noindent This apparent rotation of the thermal ejecta between day 126 and day 615 was due to the outflow being faster along the east--west axis, with the result that the east--west lobes became optically thin first. Just such an asymmetry is predicted by hydrodynamic simulations of interacting winds shaped by orbital evolution\cite{Soker89}. In this scenario, binary stars orbiting within the nova envelope transfer some of their orbital energy to the surrounding material through viscous interaction, thereby expelling the envelope preferentially along the orbital plane\cite{Livio90, Lloyd97} (Fig.\ 3a), corresponding to a north--south orientation in V959~Mon. This equatorial material is observable as thermal ejecta, but it expands relatively slowly, and its compact structure therefore proves difficult to image at early times. Meanwhile, a fast prolonged wind is blown off the white dwarf\cite{Kato94}, and this thermal wind preferentially expands along the low-density polar directions\cite{Porter98} (Fig.\ 3b). At early times, while the ejecta are optically thick, this fast material expanding along the poles will dominate the radio images, as in Fig.\ 2c. Later, when the thermal radio emission becomes optically thin, the dense material in the orbital plane will be brightest (Figs.\ 2d and 3c). A similar 90 degree flip of the major axis has been hinted at in radio imaging of other novae\cite{Taylor88, Eyres92, Heywood05}, suggesting that such a transformation may be common in classical novae.

\noindent The VLBI knots, and by extension the $\gamma$-ray emission, appear to be produced in the interaction between the rapidly expanding material driven along the poles and the slower equatorial material (Fig.\ 3c). This interaction within the ejecta could explain the prolonged duration of the radio synchrotron emission\cite{Shankar91}, lasting as long as the fast wind flows past the dense material concentrated in the orbital plane. 

\noindent The mass ejection observed in V959~Mon is a version of the common envelope phase that occurs in all close binary stars, and is a critical step in the formation of diverse phenomena like X-ray binaries, Type Ia supernovae, and stripped-envelope supernovae. Despite its widespread significance, common envelope evolution remains one of the most poorly understood phases of binary evolution, with few observational tests and models that often fail to expel the envelope at all\cite{Passy12, Ivanova13}. V959~Mon shows that classical novae can serve as a testbed for developing an understanding of common envelope evolution, and that common-envelope interaction plays a role in ejecting nova envelopes.

\noindent An extensive multi-wavelength observational campaign shows V959~Mon to be a typical classical nova. Its expansion velocities, spectral line profiles, binary period, binary companion, and optical light curve fall well within expected ranges\cite{Munari13, Ribeiro13, Shore13, Page13}. Additionally, after the few early epochs showing a flat radio spectrum, the radio light curve of V959~Mon became consistent with thermal emission from expanding warm ejecta, implying $4\times10^{-5}$ M$_{\odot}$ of ejected mass (a typical value for a classical nova\cite{Gehrz98}; Extended Data Fig.\ 4; Supplementary Information). The only unusual characteristic of V959 Mon is its proximity; at a distance of $\ltrsim$2 kpc\cite{Munari13}, it is several times closer than the typical nova which explodes in the Galactic bulge (d $\approx$ 8 kpc). Therefore, $\gamma$-rays could be a common feature of normal classical novae.

\noindent Since the outburst of V959~Mon, three additional classical novae have been identified with \emph{Fermi}\cite{Cheung13, Cheung13a}, further implying that V959~Mon is not unusual, and many novae produce $\gamma$-rays. The recent uptick in \emph{Fermi} detections of novae can likely be explained by a combination of deeper, targeted detection efforts and a lucky crop of nearby novae; with effort, the sample of $\gamma$-ray-detected novae will continue to grow. The mechanism we propose here for powering the $\gamma$-ray emission in V959~Mon---binary interaction shaping nova ejecta and powering strong internal shocks---works in most novae, implying that each of these garden-variety explosions accelerates particles to relativistic speeds.



\begin{addendum}

\item[Supplementary Information] is linked to the online version of the paper at www.nature.com/nature.

\item National Radio Astronomy Observatory (NRAO) is a facility of the National Science Foundation (NSF) operated under cooperative agreement by Associated Universities, Inc. The EVN is a joint facility of European, Chinese, South African, and other radio astronomy institutes funded by their national research councils. The EVN and e-VLBI research infrastructures were supported by the European Commission Seventh Framework Programme (FP/2007-2013) under grant agreements \#283393 (RadioNet3) and \#RI-261525 (NEXPReS). e-MERLIN is operated by The University of Manchester at Jodrell Bank Observatory on behalf of the Science and Technology Facilities Council.  The SMA is a joint project between the Smithsonian Astrophysical Observatory and the Academia Sinica Institute of Astronomy and Astrophysics.
Support for CARMA construction was derived from the 
Moore Foundation, 
Norris Foundation, 
McDonnell Foundation, the Associates of the California Institute of Technology, the University of Chicago, the states of California, Illinois, and Maryland, and the NSF. Ongoing CARMA development and operations are supported by the NSF
and by the CARMA partner universities. L.C. is a Jansky Fellow of NRAO. This research has received funding from NASA programs DPR S-15633-Y and 10-FERMI10-C4-0060 (C.C.C.), NASA award NNX13AO91G (T.N.), NSF award AST-1211778 (J.L.S. and J.W.), the South African SKA Project (V.A.R.M.R.), and  the Alexander von Humboldt Foundation (N.R.).

\item[Author Contributions] L.C. wrote the text. L.C., J.L., J.Y., T.J.O., Z.P., A.J.M., C.C.C., R.J.B, T.N., Y.Z., J.W., and G.B.T. obtained and reduced the data. All authors contributed to the interpretation of the data and commented on the final manuscript.

\item[Competing Interests] The authors declare that they have no competing financial interests.

\item[Correspondence] Correspondence and requests for materials should be addressed to L.C. (email: chomiuk@pa.msu.edu).

\end{addendum}

\clearpage

\bigskip \noindent {\bf Fig.\~1: Radio light curves and spectra of V959~Mon. Panel (a):} The multi-frequency light curve (Extended Data Tables 1 and 2) is as expected for expanding thermal ejecta, except at the earliest times ($<$30 days) and lowest frequencies ($<$2 GHz; see Supplementary Information). The time of optical identification is marked (dashed line). The times corresponding to three select radio spectra are marked with grey bars. Panel \textbf{(b)} shows an early-time flat spectrum (16 days after \emph{Fermi} discovery), transitioning to \textbf{(c)} an optically-thick thermal spectrum (day 145), and \textbf{(d)} a late thermal spectrum that is mostly optically thin (day 542). Error bars denote 1$\sigma$ uncertainty. 

\bigskip \noindent {\bf Fig.\~2: Radio imaging of V959~Mon. Panel (a)} illustrates the expansion of the compact radio knots, with a high-resolution 5 GHz EVN image from 113 days after $\gamma$-ray discovery shown in color, and contours representing the EVN image from day 91. Contour levels span 0.125--2 mJy/beam in steps of 0.125 mJy/beam. \textbf{Panel (b)} shows the 5.8 GHz e-MERLIN color image of thermal nova ejecta on day 87. The compact VLBI knots from day 91 are superimposed as contours, with levels as in Panel (a). \textbf{Panel (c)} shows similar components as Panel (b), but one month later, with a 36.5 GHz VLA color image of the the thermal ejecta on day 126. The VLBI knots from day 113 are contours. \textbf{Panel (d)} shows the expansion and flip of the thermal nova ejecta, comparing VLA images from four months and two years after outburst. The 17.5 GHz image from day 615 is shown in color, overplotted with the 36.5 GHz day 126 image now in black contours (levels 3.2, 9.6, 16, 22.4, and 28.8 mJy/beam).  In all panels, the presumed location of the binary is marked as a white cross. Scale bars in Panels (b) and (d) assume a distance of 1.5 kpc\cite{Munari13}. Synthesized beams for contours are shown in the bottom left corners of Panels (a) and (d). White date labels are for color images; gray labels for contours. 

\bigskip \noindent {\bf Fig.\~3:~Simple illustration of the 2012 outburst of V959~Mon. Panel (a)} Immediately following the thermonuclear runaway, the nova envelope expands (yellow ellipse) and interacts with the binary, yielding dense material in the equatorial plane (darker yellow; here oriented vertically)\cite{Livio90}. Denser bow shocks surround this puffy equatorial disk\cite{Porter98} (orange lines). \textbf{Panel (b)} As the nova outburst progresses, the white dwarf powers a fast wind\cite{Kato94, Lloyd97}, funneled toward the low-density poles\cite{Soker89, Porter98} (blue cones; compare with the thermal emission imaged with VLA on day 126; see Fig.\ 2c). The differential velocity produces shocks\cite{Shankar91} (orange lines), and at the edges of the optically thick ejecta, the shocked material yields compact radio knots (red blobs; compare with VLBI knots in Fig.\ 2). \textbf{Panel (c)} Once the white dwarf wind ceases, the polar outflow will detach from the binary and quickly drop in density as it expands (blue cones). The slower-expanding equatorial material will remain dense for longer (yellow regions), and will dominate the radio images at late times (compare with day 615 VLA imaging; Fig.\ 2d).

\begin{figure}
\includegraphics[width=5.8in]{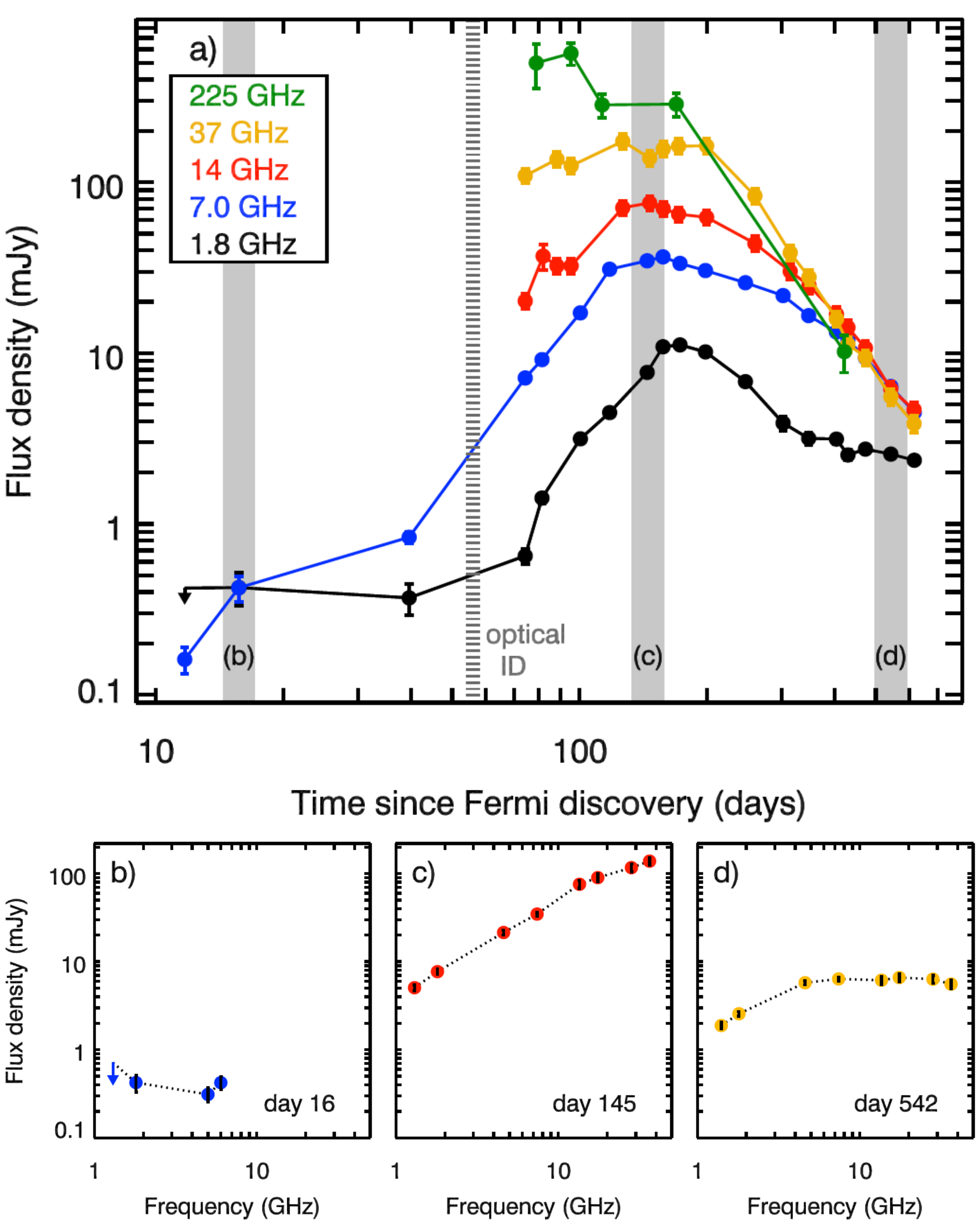}
\end{figure}

\begin{figure}
\centering
\includegraphics[width=6.2in, angle=0]{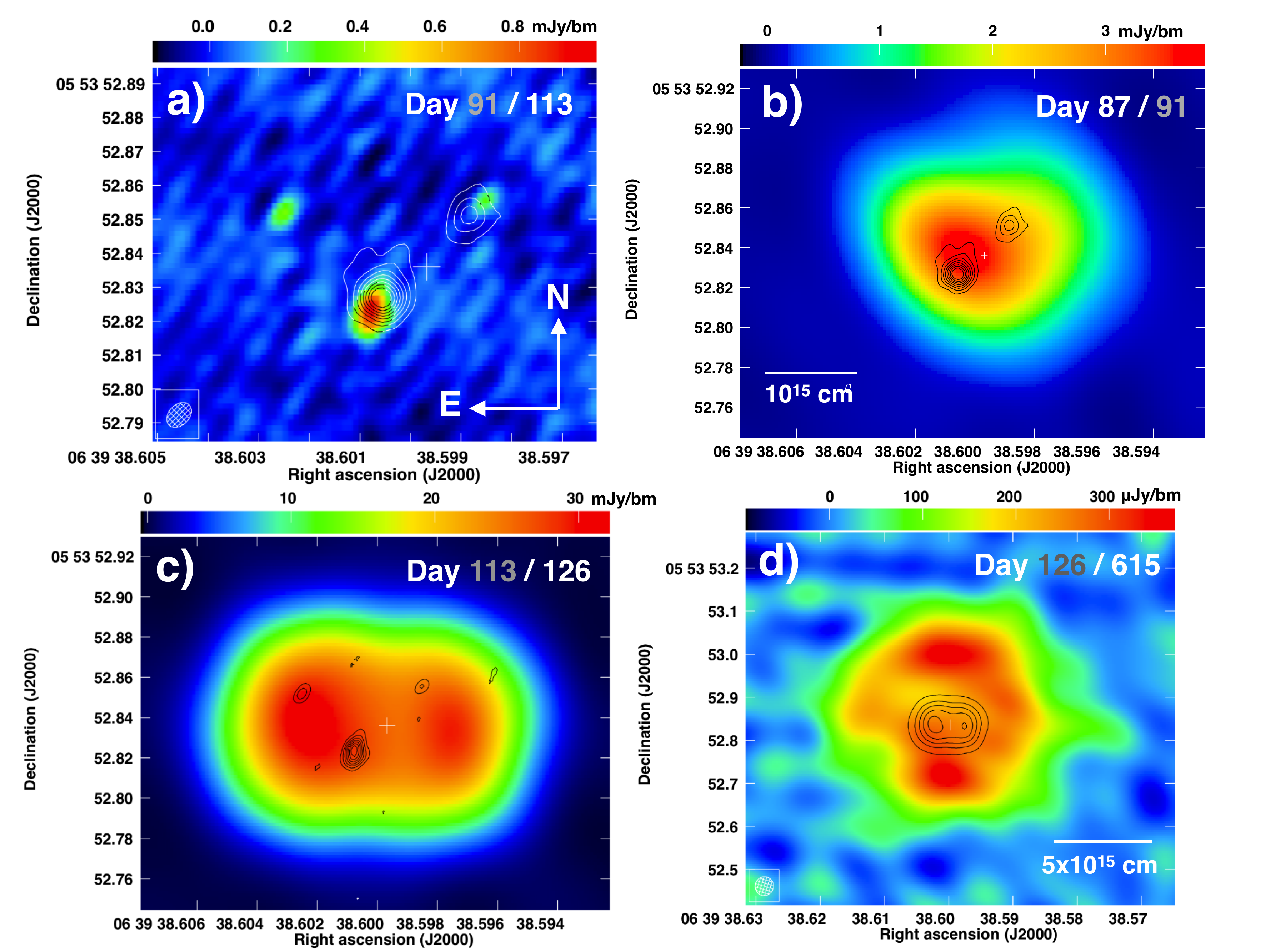}
\end{figure}

\begin{figure}
\centering
\includegraphics[width=4in, angle=0]{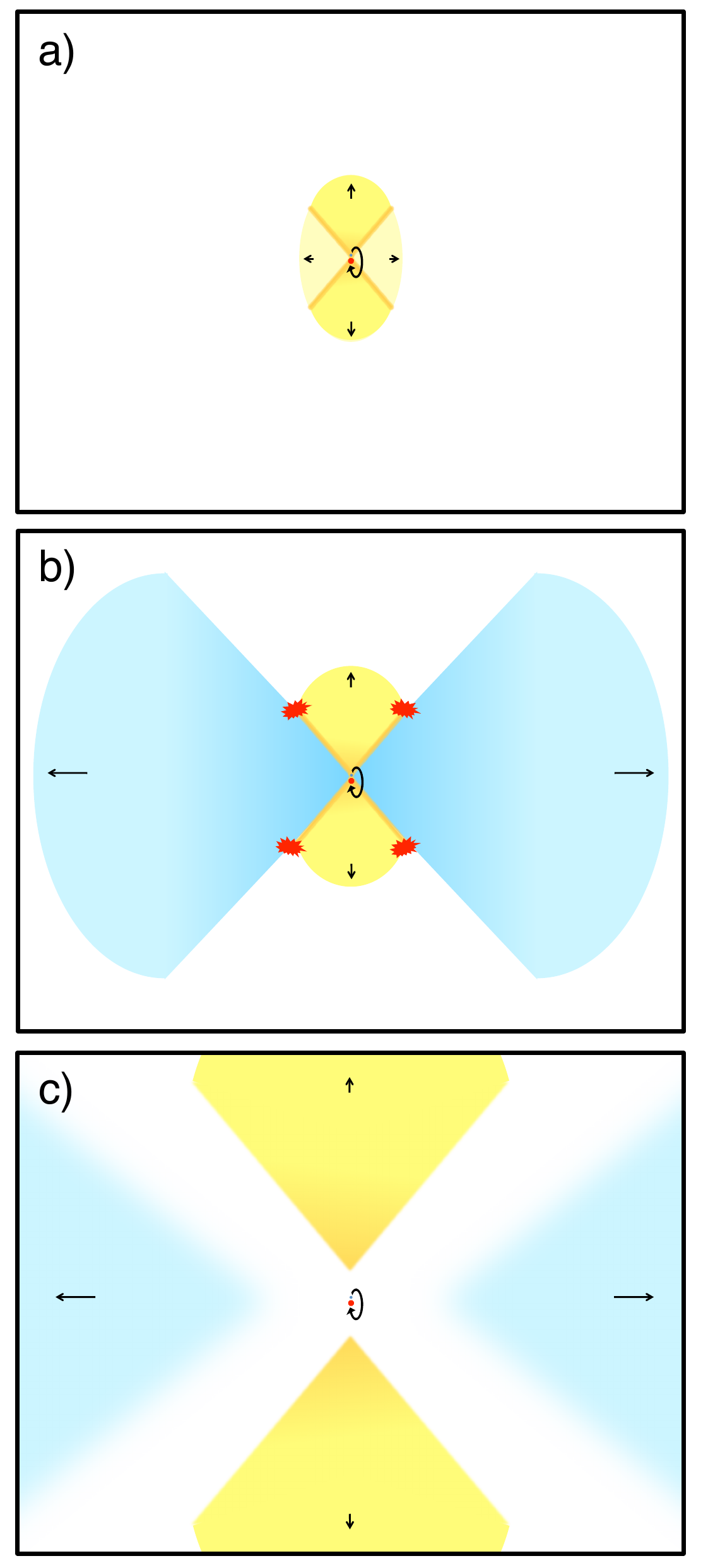}
\end{figure}



\clearpage
\noindent {\bf {\Large Methods}}\\

\section*{VLA} 
We observed V959~Mon with the VLA from Jun 2012\cite{Chomiuk12} until the time of publication as part of programs S4322, 12B-375, 13A-461, 13B-057, and S61420. Monitoring was carried out at L (1--2 GHz), C (4--8 GHz), Ku (12--18 GHz), and Ka (26--36 GHz) bands across all array configurations\cite{Napier83}.  
In C, Ku, and Ka bands, we observed with two sidebands, each composed of 8 spectral windows with 64 2-MHz-wide channels and four polarization products. Each sideband had a bandwidth of 1 GHz, and the two sidebands were separated to enhance our frequency coverage. In L band, we only obtained 1 GHz of bandwidth total, covering the entire available frequency range in this band. 

In the first three epochs, the entire positional error circle of the \emph{Fermi} transient was observed (95\% confidence radius of 11 arcmin), with a single pointing at L band and a seven-pointing mosaic at C band. Later, after identification of the optical nova, a single pointing was centered on V959~Mon, typically yielding $\sim$10--15 minutes on source in each band and epoch. 

We observed J0632+1022 as the secondary phase calibrator at L and C bands, and J0643+0857 at Ku and Ka bands. 3C147 was used as an absolute flux density and bandpass calibrator. Data were reduced using standard routines in CASA and AIPS, and in most cases, a single round of phase-only  self calibration was implemented with 1-minute solution intervals. Imaging was carried out in CASA\cite{McMullin07}, AIPS\cite{Greisen03}, and Difmap\cite{Shepherd94}, typically using a Briggs robust value\cite{Briggs95} of 1. Flux densities were obtained by fitting a gaussian to the source, using \verb|JMFIT| in AIPS or \verb|gaussfit| in CASA. Measurements are presented in Table ED1. Uncertainties include 5\% calibration error at L and C bands and 10\% calibration error at Ku and Ka bands.

In the A configuration epochs, which provide the highest angular resolution, special attention was devoted to imaging. To achieve the highest possible resolution, the VLA images shown in Fig.\ 2 were produced in Difmap with uniform weighting. The 36.5 GHz image from 2012 Oct 23 featured in Fig.\ 2c has a synthesized beam of $44 \times 42$ mas full width at half maximum (FWHM) and an rms sensitivity of 73 $\mu$Jy/beam. The 17.5 GHz image from 2014 Feb 24 in Fig.\ 2d has a $0.125^{\prime\prime} \times 0.09^{\prime\prime}$ FWHM synthesized beam and rms noise of 19 $\mu$Jy/beam. When smoothed to similar resolution, images at other frequencies show similar structure on this date. We note that during the first A configuration (Oct 2012--Jan 2013), the nova was not spatially resolved at the lower frequencies (L and C bands). In the second A configuration (Feb 2014), V959~Mon was slightly resolved at C band and remained unresolved at L band.

\section*{Millimeter Data} 
The 225 GHz flux density of V959~Mon was monitored using the Submillimeter Array (SMA) and Combined Array for Research in Millimeter-wave Astronomy (CARMA). Under SMA program 2012A-S016, observations were obtained from Sep--Dec 2012. CARMA observations were obtained at 96 and 225 GHz during May and Aug 2013 under program c1130. Data were reduced using standard routines in MIRIAD\cite{Sault95}, and measurements are listed in Table ED2. Uncertainties include calibration errors of at least 10\%.

In addition, we utilized high-frequency measurements obtained with the Institute for Radio Astronomy in the Millimeter Range (IRAM) 30m telescope and Plateau de Bure Interferometer (PdBI)\cite{Fuhrmann12} (Table ED2). These measurements additionally constrained our radio/millimeter spectrum fitting and light curve modeling, as shown in Figs.\ ED1 and ED4.

\section*{e-MERLIN} The e-MERLIN array was used to observe V959 Mon on a number of occasions starting Sep 2012, when the VLA detected an increased radio brightness. The first epoch (shown in Fig.\ 2b) combined observations from 2012 Sep 13 and 14. The observations used six telescopes of the e-MERLIN array. They were made at C-band, centered at 5.75 GHz with a bandwidth of 512 MHz per polarization. This total bandwidth was split into four adjacent 128-MHz-wide sub-bands and correlated with 512 channels per sub-band. We observed J0645+0541 as a phase calibrator, 3C286 as a flux calibrator and OQ208 as a bandpass calibrator. Data were calibrated and reduced with standard tasks in AIPS. The data were phase self-calibrated and imaging was carried out in AIPS, using a Briggs robust parameter of unity. The synthesized beam used to restore the image is $65 \times 45$ mas. The peak flux density in this map is 3.7 mJy/beam, with rms noise of 53 $\mu$Jy/beam. The total flux density is estimated as $7.0\pm0.5$ mJy (obtained via Gaussian fitting to the image data), which is consistent with estimates made from VLA observations on shorter baselines around the same time.

\section*{EVN} 

We performed the EVN observations of V959~Mon at C-band (5.0 GHz) in five epochs spanning Sep 2012 to Jan 2013. Observations were carried out with a data rate of 1024 Mbit s$^{-1}$ and 2-bit sampling, yielding dual polarization and 128 MHz bandwidth. Observation duration was $\sim$7 hours per epoch.  The first three observations were EVN Target of Opportunity experiments during the e-EVN sessions (Project Codes RO005 and RO006), while the last two epochs were part of project EO011. Participating EVN stations were Effelsberg, the phased array of the Westerbork Synthesis Radio Telescope, Onsala, Medicina, Noto, Torun, and the Lovell telescope, except Effelsberg was not functioning during the last epoch due to snow. Through the wide-band internet connection to the EVN stations, all observations were correlated in real time at the Joint Institute for VLBI in Europe (JIVE) with the default correlation parameters: 2 second integration time and 32 frequency points per sub-band.

We observed J0645+0541 as the phase-referencing source in all our epochs. The J2000 coordinates were RA = $06^{h}45^{m}47^{s}.27653$ and Dec = $05^{\circ}41^{\prime}22^{\prime\prime}.3857$ (absolute 1$\sigma$ positional uncertainty: 1.1 mas in RA and 2.0 mas in Dec). The separation between the nodding calibrator and V959~Mon was $1^{\circ}.54$. We also observed a secondary calibrator as an astrometric check source, which implied a 1$\sigma$ systematic position error of 0.43 mas in RA and 0.70 mas in Dec. We took a nodding cycle time of 1 minute on the calibration and 3 minutes on the target.

Initial data calibration was carried out in AIPS using standard routines. After this calibration, the data were averaged in each sub-band. Imaging and deconvolution were then carried out in Difmap. To remove the phase error associated with the source structure, we first imaged the reference source J0645+0541 and then re-did fringe-fitting with the calibrator image. The source J0645+0541 showed a single side core-jet structure with a total flux density $175\pm18$ mJy. The centroid of the radio core is determined by Gaussian model fitting and later used as the image reference origin. We imaged V959~Mon with natural weighting in Difmap. 

The intensity image in the first epoch (2012 Sep 18) had a synthesized beam with FWHM $9.2\times5.5$ mas at a position angle $-38^{\circ}.0$ and rms sensitivity of 0.03 mJy/beam. A pair of knots were clearly detected\cite{O'Brien12} (Table ED3). Here we name the brighter (eastern) one as A and the weaker (western) one as B; this image is shown as white contours in the top panel of Fig.\ 2. The two knots A and B had an angular separation of $35.5\pm0.2$ mas. After fitting the knots with point source models, there still existed some extended emission with a peak brightness of 0.27 mJy/beam in the residual map. We modeled the residual extended emission with two circular Gaussian components (FWHM: 16.4 and 29.9 mas). Each knot was associated with one extended emission region. 

We detected the pair of radio knots again in the second epoch (2012 Oct 10), along with a new, third knot (Fig.\ 2a). Knots A and B had faded since the first epoch (Table ED3). Knot A was resolved and its intensity distribution was well fit by two point sources with a separation of $5.7\pm0.2$ mas at a position angle $-179^{\circ}.5$. The separation was $45.2\pm0.4$ mas at a position angle $131^{\circ}.9$ between A1 and B and $49.1\pm0.4$ mas at position angle $137^{\circ}.0$ between A2 and B. The new third knot, dubbed C, was located east of A, and was brighter than knot B in this epoch. After fitting and removing the knots, there was some residual large-scale structure showing as regular stripes, most likely a hint of the extended emission seen with the VLA. The synthesized beam had a FWHM $8.4\times6.0$ mas at a position angle $-44^{\circ}.3$. 

We find that both radio knots A and B appeared to be expanding away from a central position. Between 2012 Sep 18 and 2012 Oct 10, they were measured to travel apart from one another at 0.50 mas/day.

We failed to detect any compact features associated with V959~Mon in the following three EVN epochs (2012 Nov 14, 2012 Dec 4, and 2013 Jan 15) because of its low peak brightness and significant expansion. We place $5\sigma$ upper limits on the peak flux density from compact knots of $<$0.25 mJy/beam on 2012 Nov 14, $<$0.15 mJy/beam on 2012 Dec 4, and $<$0.18 mJy/beam on 2013 Jan 15. There existed a hint of extended emission in the dirty map, but it was not possible to locate the position due to strong side lobes of the synthesized beam. The source was only weakly detected on the shortest and the most sensitive baseline (Effelsberg--Westerbork) in the third and fourth epochs. V959~Mon was completely buried under the noise in the last EVN epoch.

Positions and fluxes of the EVN components are listed in Table ED3.

\section*{VLBA} 
We observed V959~Mon with the VLBA on 2012 Oct 3, Oct 14, Oct 30, and Nov 17 under the NRAO project code BM0385. For the first three epochs, we observed at both L and C bands, while the final epoch was in C band. Each observation had eight spectral windows covering a total bandwidth of 256 MHz via the new ROACH digital backend (RDBE) and yielded total on-source time of $\sim$100 minutes per frequency band. For the first epoch, each spectral window had eight channels with 4 MHz bandwidth per channel. For the subsequent observations, we had 64 channels per spectral window with 500 kHz per channel. The first three epochs had center frequencies of 1.55 GHz and 4.98 GHz for L and C band, respectively. The final C band epoch had a center frequency of 4.24 GHz. The C band phase reference source was J0650+0358 for the first two epochs and J0645+0541 for the final two epochs (changed to match the calibrator used for EVN observations). The L-band phase reference source was J0650+0358 for all epochs. The data were reduced using standard routines in AIPS, and images were made using both AIPS and Difmap.

In our first VLBA epoch, we detected VLBI knots A and B in both the 5.0 GHz and 1.6 GHz data.  For the 5.0 GHz image ($3.5 \times 1.5$ mas synthesized beam; rms sensitivity 27 $\mu$Jy/beam), knot A was not well fit with a point source and was therefore modeled with a circular Gaussian, while knot B was well fit by a point source model (Table ED3).  We also noticed a possible third radio knot approximately 7.7 mas due north of the brightest component.  For the 1.6 GHz image ($11 \times 5$ mas synthesized beam; rms sensitivity 49 $\mu$Jy/beam), both components A and B were diffuse and modeled with circular Gaussians.  There was no sign of the third component in the 1.6 GHz image.

For the second VLBA epoch, we detected the three EVN knots in the 5.0 GHz image ($3.8 \times 1.7$ mas synthesized beam; rms sensitivity 37 $\mu$Jy/beam).  Knots A and B had faded with respect to the first VLBA observation, and knot C was seen to the east of A.  Knot A was best fit by two point source models and a circular Gaussian component to account for diffuse emission.  Knots B and C were well fit by single point sources.  In the 1.6 GHz data ($12 \times 6$ mas synthesized beam; rms sensitivity 27 $\mu$Jy/beam), only knot B was obvious.  It was again fit with a circular Gaussian.  There was a second small knot in the vicinity of A, and this was also fit with a circular Gaussian, but we do not believe it accounts for the total flux in knot A.

Unfortunately, the phase reference source did not produce strong fringes on the longer baselines at 5.0 GHz for the third and fourth VLBA epochs.  In addition, one antenna was down during each of these observations.  The reduced sensitivity and lower resolution led to problems identifying distinct components, while the change in phase reference source makes it difficult to compare positions between the first and last two epochs.  For the third epoch at 5.0 GHz, knot A had a flux density of $1.0\pm0.1$ mJy, while knot B was not detected with high confidence (upper limit of $0.14$ mJy).  At 1.6 GHz, only one component in the vicinity of knot A was detected unambiguously, but with a dramatically increased flux density of $1.2\pm0.1$ mJy.  Knot B had an upper limit of $0.14$ mJy.  The increase in flux density for knot A and drop in flux density for B could indicate the shock associated with knot A encountered a region of higher density leading to increased synchrotron emission, while knot B exhausted its supply of relativistic particles.  For the fourth epoch, no 1.6 GHz data were obtained.  The source had dimmed significantly at 5.0 GHz, and knot A was just barely detected with a flux density of $0.16\pm0.05$ mJy.  Knot B was not detected (upper limit of 0.11 mJy).

Positions and fluxes of the VLBA components are listed in Table ED3. Note that some small disagreement between EVN and VLBA positions is expected due to the use of different phase reference sources.

Concurrent observations at 1.6 and 5.0 GHz enable creation of a spectral index map.  For the 2012 Oct 3 observations, we applied a \textit{(u,v)}-taper to the 5.0 GHz data to match resolution with our 1.6 GHz image.  The matched resolution images were then aligned via 2-dimensional cross correlation, and low signal-to-noise regions were blanked. We used the AIPS task \verb|SPIXR| to produce the map presented in Fig.\ ED2.  Knot A had an overall positive spectral index, with a mean $\alpha$ in the high signal-to-noise region of approximately $1.2$.  Knot B had an overall negative spectral index, with a mean value of $\alpha \approx -0.3$ in the regions of highest signal-to-noise.  The spectra were generally flatter ($\alpha$ closer to $0$) near the edges of the knots. 

The spectral index of Knot B appears roughly constant between days 106 and 117 (Table ED3). Curiously, Knot A appeared to have faded at low frequency on day 117 (leading to $\alpha > 1.2$), but then brightened dramatically at 1.6 GHz on day 133 (yielding $\alpha \approx 0$). The optical depth of Knot A therefore appears to be time variable.

From our two epochs of 5 GHz data, we can get an independent estimate of the expansion rate for knots A and B.  As stated previously, the EVN observations exhibited an expansion rate of $\sim0.50$ mas/day between 2012 Sep 18--Oct 10. Using the VLBA positions listed in Table ED3, we find an expansion rate of $\sim0.35$ mas/day between 2012 Oct 3--Oct 14. Because the VLBA observations were made later than the EVN, the slower rate may indicate that the expansion was slowing down.  However, possible evolution of the source structure and the different \textit{(u,v)} coverages of the instruments also introduce significant uncertainties into this comparison. Because we only have two epochs with detections in each array, it proves difficult to establish the detailed evolution of the VLBI knot expansion velocity.

\section*{Expansion of the Thermal Ejecta: Days 127--199}

During the first period of VLA A configuration, we obtained five epochs spanning 2012 Oct 23--2013 Jan 4 (Table ED1). V959 Mon was clearly resolved at frequencies 13.5--36.5 GHz. V959~Mon was bright during this period, and imaged at significance levels of $>$100$\sigma$ per synthesized beam. In each image, V959~Mon was fit with a gaussian using the task \verb|JMFIT| in AIPS. The width and position angle of the gaussian were allowed to vary, and the angular dimensions of V959~Mon were found by deconvolving the synthesized beam from the fitted gaussian. 

Gaussian fits provide simple first-order estimates of the changing dimensions of V959~Mon, although the source is not perfectly described with this profile form (future work will involve a more detailed analysis of the source geometry). The position angles of the fitted gaussians were constant with time, consistent across frequency ($\sim87^{\circ}$), and distinct from the position angle of the synthesized beams. V959~Mon obviously expanded along both its major (east--west) and minor (north--south) axes over the three months of A configuration observations. Fig.\ ED3 shows the deconvolved radii of V959~Mon over this period, with the semi-major axis plotted in the top panel and semi-minor axis in the bottom panel. Both axes were observed to increase with remarkable linearity at all four frequencies; linear fits are overplotted. At 13.5 GHz, we find diameter expansion rates of 0.62 mas/day and 0.64 mas/day along the minor and major axes, respectively. At 36.5 GHz, we find diameter expansion rates of 0.66 mas/day along the minor axis and 0.37 mas/day for the major axis.

At any given epoch, very similar minor-axis radii were measured at all four frequencies. However, the size of the major axis varied with frequency, with the exception of the first epoch. 36.5 GHz measurements of the major axis yielded systematically smaller radii than 13.5 GHz measurements (and data points at the intermediate frequencies of 17.5 and 28.2 GHz also follow this trend). Assessing all five epochs, it is clear that the major axis of V959~Mon grew most slowly at highest frequency.

The material expanding along V959~Mon's minor axis was consistent with being optically thick throughout the A configuration observations. When optically thick, all frequencies display roughly the same radius of the $\tau \approx 1$ surface, as we observed for material expanding along the minor axis of V959~Mon. On the other hand, the frequency dependence of V959~Mon's major axis can be explained if the ejecta expanding along the major axis were becoming optically thin over the course of the imaging campaign.




\newpage

\bigskip \noindent {\bf Table~ED1:~VLA Observations of V959~Mon.} Flux density measurements at 1--46 GHz, spanning Jun 2012--Feb 2014. The time $t_0$ corresponds to $\gamma$-ray discovery, 2012 June 19\cite{Cheung13}.

\bigskip \noindent {\bf Table~ED2:~Millimeter Observations of V959~Mon.} Flux density measurements obtained with the SMA, CARMA, IRAM, and PdBI telescopes at 86--230 GHz. The time $t_0$ corresponds to $\gamma$-ray discovery, 2012 June 19\cite{Cheung13}.

\bigskip \noindent {\bf Table~ED3:~VLBI Components of V959~Mon.} Positions and flux densities of VLBI knots, observed with EVN and VLBA. The time $t_0$ corresponds to $\gamma$-ray discovery, 2012 June 19\cite{Cheung13}.

\bigskip \noindent {\bf Fig.\~ED1:~Radio/millimeter spectral evolution of V959~Mon.} Measurements and 1$\sigma$ uncertainties from select epochs are shown as black points. Power law or broken power law fits are overplotted as red lines (the function is chosen to minimize the reduced $\chi^2$ value). The best-fit spectral indices are listed in each panel, along with the break frequency ($\nu_b$) in the case of broken power law fits. Errors correspond to 1$\sigma$ uncertainty.

\bigskip \noindent {\bf Fig.\~ED2:~Spectral index map from 2012 October 3 VLBA observations.}  The spectral index is measured by comparing images at 1.6 and 5 GHz. Overlaid contours are from 1.6 GHz Stokes I map. Contour levels are -0.08, 0.08, 0.13, 0.16, 0.23, 0.32, and 0.45 mJy/beam.

\bigskip \noindent {\bf Fig.\~ED3:~The expansion of V959~Mon as a function of time. Panel (a)} plots semi-major axis, and {\bf Panel (b)} plots semi-minor axis, both in units of milli-arcseconds. Measurements at four distinct frequencies are plotted in different colors, as defined in the key at upper left. Error bars from JMFIT are so small that they are not visible. Linear fits are made to each frequency separately, and are plotted as colored lines. 

\bigskip \noindent {\bf Fig.\~ED4:~Model fit to the radio/millimeter light curve of V959~Mon.} A simple model of thermal expanding ejecta roughly describes the light curve evolution at day $\sim$200 and later, and implies an ejected mass of few $\times 10^{-5}$ M$_{\odot}$. Error bars denote 1$\sigma$ uncertainty.



\pagestyle{empty}

\clearpage

\begin{deluxetable}{cccccccccccccc}
\captionsetup[table]{labelformat=empty, labelsep=none}
\tablewidth{0 pt}
\tabletypesize{\tiny}
\setlength{\tabcolsep}{0.025in}
\tablehead{Date & $t-t_{0}$   & Conf. & $\nu$ & $S_\nu$ & $\nu$ & $S_\nu$ & $\nu$ & $S_\nu$ & $\nu$ & $S_\nu$ \\
(UT) & (Days) & & (GHz) & (mJy) & (GHz) & (mJy) & (GHz) & (mJy) & (GHz) & (mJy)}
\startdata

2012 Jun 30.7 & 11.7 & B & 1.3& $  0.10\pm 0.18$ & 1.8& $  0.15\pm 0.09$ & 5.0& $  0.26\pm 0.03$ & 6.0& $  0.16\pm 0.03$ \\
 &  &  &  & & & & & & & & \\

2012 Jul 4.7 & 15.7 & B & 1.3& $  0.11\pm 0.20$ & 1.8& $  0.42\pm 0.09$ & 5.0& $  0.31\pm 0.06$ & 6.0& $  0.42\pm 0.07$ \\
 &  &  &  & & & & & & & & \\

2012 Jul 28.6 & 39.6 & B & 1.4& $  0.54\pm 0.14$ & 1.8& $  0.37\pm 0.08$ & 5.0& $  0.70\pm 0.06$ & 6.0& $  0.84\pm 0.07$ \\
 &  &  &  & & & & & & & & \\

2012 Sep 1.4 & 74.4 & A & 1.3& $  0.39\pm 0.07$ & 1.8& $  0.65\pm 0.07$ & 4.6& $  3.52\pm 0.18$ & 7.4& $  7.16\pm 0.36$ \\
& & & 13.3& $ 20.20\pm 2.02$ & 17.4& $ 30.82\pm 3.08$ & 27.5& $ 73.59\pm 7.36$ & 36.5& $109.3\pm10.9$ \\
 &  &  &  & & & & & & & & \\

2012 Sep 8.5 & 81.5 & BnA &  1.3& $  1.24\pm 0.10$ & 1.8& $  1.42\pm 0.10$ & 4.6& $  4.74\pm 0.24$ & 7.4& $  9.16\pm 0.46$ \\
 &  &  &  & & & & & & & & \\

2012 Sep 15.4 & 88.4 & BnA & 13.5& $ 32.46\pm 3.25$ &17.4& $ 47.25\pm 4.73$ &28.2& $102.6\pm10.3$ & 36.5& $136.6\pm13.7$ \\
& & & 40.6& $211.7\pm21.2$ & 45.4& $235.8\pm23.6$ & &\\
 &  &  &  & & & & & & & & \\

2012 Sep 22.5 & 95.5 & BnA & 13.5& $ 32.44\pm 3.26$ &17.4& $ 61.76\pm 6.25$ &28.5& $ 96.15\pm 9.66$ & 36.5& $124.9\pm12.6$\\
 &  &  &  & & & & & & & & \\

2012 Sep 27.4 & 100.4 & BnA$\rightarrow$A &  1.3& $  2.55\pm 0.14$ & 1.8& $  3.15\pm 0.17$ & 4.6& $  9.00\pm 0.45$ & 7.4& $ 17.21\pm 0.88$ \\
 &  &  &  & & & & & & & & \\

2012 Oct 14.7 & 117.7 & A &  1.3& $  2.75\pm 0.16$ & 1.7& $  4.49\pm 0.24$ & 4.6& $ 16.28\pm 0.82$ & 7.4& $ 30.99\pm 1.57$\\
 &  &  &  & & & & & & & & \\

2012 Oct 23.4 & 126.4 & A & 13.6& $ 71.03\pm 7.10$ &17.5& $ 89.70\pm 8.97$ &28.2& $138.9\pm13.9$ & 36.5& $173.9\pm17.4$ \\
 &  &  &  & & & & & & & & \\

2012 Nov 10.3 & 144.3 & A &  1.3& $  5.06\pm 0.38$ & 1.8& $  7.71\pm 0.48$ & 4.6& $ 21.45\pm 1.36$ & 7.4& $ 34.73\pm 1.78$ \\
 &  &  &  & & & & & & & & \\

2012 Nov 12.3 & 146.3 & A & 13.5& $ 75.70\pm 7.60$ &17.5& $ 90.08\pm 9.03$ &28.2& $116.9\pm11.9$ &36.5& $138.6\pm14.0$\\
 &  &  &  & & & & & & & & \\

2012 Nov 23.4 & 157.4 & A &  1.3& $  7.61\pm 0.39$ & 1.8& $ 10.90\pm 0.55$ & 4.6& $ 25.94\pm 1.30$ & 7.4& $ 36.60\pm 1.83$ \\
& & &13.6& $ 70.15\pm 7.06$ &17.5& $ 87.11\pm 8.74$ &28.2& $135.3\pm14.0$ &36.5& $156.9\pm16.1$ \\
 &  &  &  & & & & & & & & \\

2012 Dec 7.2 & 171.2 & A &13.5& $ 65.08\pm 6.51$ &17.5& $ 80.92\pm 8.09$ &28.2& $133.5\pm13.4$ & 36.5& $163.0\pm16.3$ \\
 &  &  &  & & & & & & & & \\

2012 Dec 8.5 & 172.5 & A &  1.3& $  8.40\pm 0.43$ & 1.8& $ 11.17\pm 0.56$ & 4.6& $ 23.76\pm 1.19$ & 7.4& $ 33.51\pm 1.68$ \\
 &  &  &  & & & & & & & & \\

2013 Jan 3.1 & 198.1 & A &  1.3& $  8.16\pm 0.41$ & 1.7& $ 10.16\pm 0.51$ & 4.6& $ 20.17\pm 1.01$ & 7.4& $ 30.45\pm 1.52$ \\
 &  &  &  & & & & & & & & \\

2013 Jan 4.3 & 199.3 & A & 13.5& $ 62.45\pm 6.25$ &17.5& $ 78.38\pm 7.84$ &28.2& $134.9\pm13.5$ &36.5& $163.4\pm16.4$ \\
 &  &  &  & & & & & & & & \\

2013 Feb 20.1 & 246.1 & D &  1.4& $  6.22\pm 0.35$ & 1.8& $  6.81\pm 0.36$ & 4.6& $ 15.75\pm 0.79$ & 7.4& $ 25.83\pm 1.29$ \\
 &  &  &  & & & & & & & & \\

2013 Mar 05.1 & 259.1 & D & 13.6& $ 43.87\pm 4.39$ &17.5& $ 54.63\pm 5.46$ &28.2& $ 74.11\pm 7.41$ &36.5& $ 83.35\pm 8.34$ \\
 &  &  &  & & & & & & & & \\

2013 Apr 17.0 & 302.0 & D &  1.4& $  2.93\pm 0.36$ & 1.8& $  3.90\pm 0.37$ & 4.7& $ 12.23\pm 0.62$ & 7.3& $ 21.73\pm 1.09$ \\
 &  &  &  & & & & & & & & \\

2013 Apr 28.9 & 313.9 & D & 13.2& $ 30.12\pm 3.01$ &17.5& $ 34.66\pm 3.47$ &28.2& $ 39.48\pm 3.95$ &36.5& $ 38.69\pm 3.87$ \\
 &  &  &  & & & & & & & & \\

2013 Jun 1.4 & 347.4 & DnC$\rightarrow$C & 1.4& $  2.69\pm 0.37$ & 1.7& $  3.17\pm 0.26$ & 4.6& $ 10.53\pm 0.53$ & 7.4& $ 16.58\pm 0.83$ \\
& & &13.2& $ 24.86\pm 2.49$ &17.5& $ 27.01\pm 2.70$ &28.2& $ 29.16\pm 2.92$ &36.5& $ 27.87\pm 2.79$ \\
 &  &  &  & & & & & & & & \\

2013 Jun 27.6 & 403.6 &C &  1.4& $  1.95\pm 0.19$ & 1.8& $  3.14\pm 0.21$ & 4.6& $  9.21\pm 0.46$ & 7.4& $ 13.30\pm 0.67$ \\
& & & 13.5& $ 16.86\pm 1.69$ &17.5& $ 17.20\pm 1.72$ &28.2& $ 16.94\pm 1.70$ &36.5& $ 15.87\pm 1.59$ \\
 &  &  &  & & & & & & & & \\

2013 Aug 22.5 & 429.5 & C &  1.4& $  1.81\pm 0.23$ & 1.8& $  2.53\pm 0.18$ & 4.6& $  8.72\pm 0.44$ & 7.4& $ 11.94\pm 0.60$\\
& & &13.5& $ 14.13\pm 1.41$ &17.5& $ 14.11\pm 1.41$ &28.2& $ 13.09\pm 1.32$ &36.5& $ 11.27\pm 1.14$ \\
 &  &  &  & & & & & & & & \\

2013 Oct 04.4 & 472.4 & B &  1.4& $  1.96\pm 0.14$ & 1.8& $  2.74\pm 0.16$ & 4.6& $  7.60\pm 0.38$ & 7.4& $  9.38\pm 0.47$ \\
& & & 13.6& $ 10.71\pm 1.07$ &17.5& $ 10.53\pm 1.05$ &28.2& $ 10.11\pm 1.01$ &36.5& $  9.45\pm 0.96$ \\
 &  &  &  & & & & & & & & \\

2013 Dec 13.2 & 542.2 & B &   1.4& $  1.88\pm 0.17$ & 1.8& $  2.56\pm 0.15$ & 4.6& $  5.76\pm 0.29$ & 7.4& $  6.38\pm 0.32$ \\
& & & 13.6& $  6.16\pm 0.62$ &17.5& $  6.60\pm 0.66$ &28.2& $  6.33\pm 0.64$ &36.5& $  5.56\pm 0.59$ \\
 &  &  &  & & & & & & & & \\

2014 Feb 24.2 & 615.2 & A & 13.6& $  4.67\pm 0.48$ &17.5& $  4.23\pm 0.43$ &28.2& $  4.39\pm 0.48$ &36.5& $  3.86\pm 0.43$ \\
 &  &  &  & & & & & & & & \\

2014 Feb 25.0 & 616.0 & A &  1.4& $  1.82\pm 0.11$ & 1.8& $  2.36\pm 0.12$ & 4.6& $  4.27\pm 0.21$ & 7.4& $  4.50\pm 0.23$ \\
 &  &  &  & & & & & & & & \\
\enddata
\end{deluxetable}

\renewcommand\thetable{ED2}
\begin{deluxetable}{cccccccccccccc}
\captionsetup{labelformat=empty}
\tablewidth{0 pt}
\tabletypesize{\tiny}
\setlength{\tabcolsep}{0.025in}
\tablehead{Date & $t-t_{0}$   & Facility & $\nu$ & $S_\nu$  \\
(UT) & (Days) & & (GHz) & (mJy) }
\startdata
2012 Sep 6 & 79 & IRAM\cite{Fuhrmann12} & 86 & $453\pm121$\\
2012 Sep 6 & 79 & SMA & 225 & $500\pm125$\\
2012 Sep 8 & 81 & PdBI\cite{Fuhrmann12} & 87 & $330\pm33$\\
2012 Sep 22 & 95 & SMA & 225 & $570\pm57$\\
2012 Oct 10 & 113 & SMA & 225 & $284\pm28$\\
2012 Dec 5 & 169 & SMA & 220 & $287\pm28$\\
2013 May 21 & 336 & CARMA & 96 & $34.7\pm3.6 $\\
2013 Aug 11 & 418 & CARMA & 96 & $13.8\pm1.5$\\
2013 Aug 14 & 421 & CARMA & 230 & $10.2\pm2.0$\\
\enddata
\end{deluxetable}

\clearpage
\renewcommand\thetable{ED3}

\begin{deluxetable}{ccccccccc}
\tablewidth{0 pt}
\captionsetup{labelformat=empty}
\tabletypesize{\tiny}
\setlength{\tabcolsep}{0.025in}
\tablehead{Date & $t-t_{0}$  & Facility & $\nu$ & $S_\nu$ & RA Pos & Dec Pos & Phase Ref.  \\
(UT) & (Days) & & (GHz) & (mJy) & (06h39mXXs) &  (05d53$^{\prime}$XX$^{\prime\prime}$) & Source}
\startdata
    \multicolumn{2}{c}{Component A}\\
\hline
 2012 Sep 18 & 91 & EVN & 5.0 & $1.04\pm0.11$ & $38.60057$ &$52.8267$ & J0645+0541\\
 2012 Oct 03 & 106 & VLBA & 1.6 & $0.53\pm0.06$ & $38.60077$ &$52.8311$ & J0650+0358\\
 2012 Oct 03 & 106 & VLBA & 5.0 & $1.66\pm0.17$ & $38.60087$ &$52.8315$ & J0650+0358\\
  2012 Oct 10 & 113  & EVN & 5.0 & $1.45\pm0.15$ & $38.60077$ &$52.8225$ & J0645+0541\\
2012 Oct 14 & 117 & VLBA & 1.6 & $0.17\pm0.05$ & $38.60054$ &$52.7915$ & J0650+0358\\
 2012 Oct 14 & 117 & VLBA & 5.0 & $1.31\pm0.14$ & $38.60097$ &$52.8327$ & J0650+0358\\
 2012 Oct 30 & 133 & VLBA &1.6 & $1.20\pm0.13$ & $38.60092$ & $52.82778$ & J0645+0541\\
 2012 Oct 30 & 133 & VLBA & 5.0 & $1.03\pm0.12$ & $38.60098$ & $52.82325$ & J0650+0358\\
 & & & & & & \\

    \multicolumn{2}{c}{Component B}\\
\hline
2012 Sep 18 & 91  & EVN & 5.0  & $0.42\pm0.05$ & $38.59886$ &$52.8514$ & J0645+0541\\
2012 Oct 03 & 106 & VLBA & 1.6 & $0.59\pm0.07$ & $38.59883$ &$52.8668$ & J0650+0358\\
2012 Oct 03 & 106 & VLBA & 5.0 & $0.16\pm0.03$ & $38.59881$ &$52.8597$ & J0650+0358\\

2012 Oct 10 & 113  & EVN & 5.0 & $0.31\pm0.04$ & $38.59852$ &$52.8555$ & J0645+0541\\
2012 Oct 14 & 117 & VLBA & 1.6 & $0.52\pm0.07$ & $38.59865$ &$52.8620$ & J0650+0358\\
2012 Oct 14 & 117 & VLBA & 5.0 & $0.16\pm0.04$ & $38.59867$ &$52.8626$ & J0650+0358\\
 & & & & & \\

    \multicolumn{2}{c}{Component C}\\
\hline
2012 Oct 10 & 113  & EVN & 5.0  & $0.36\pm0.05$ & $38.60251$ & $52.8518$ & J0645+0541\\
2012 Oct 14 & 117 & VLBA & 5.0  & $0.14\pm0.04$ & $38.60267$ & $52.8622$ & J0650+0358\\
 & & & & & \\
\enddata
\end{deluxetable}

\begin{figure}
\includegraphics[width=5.8in]{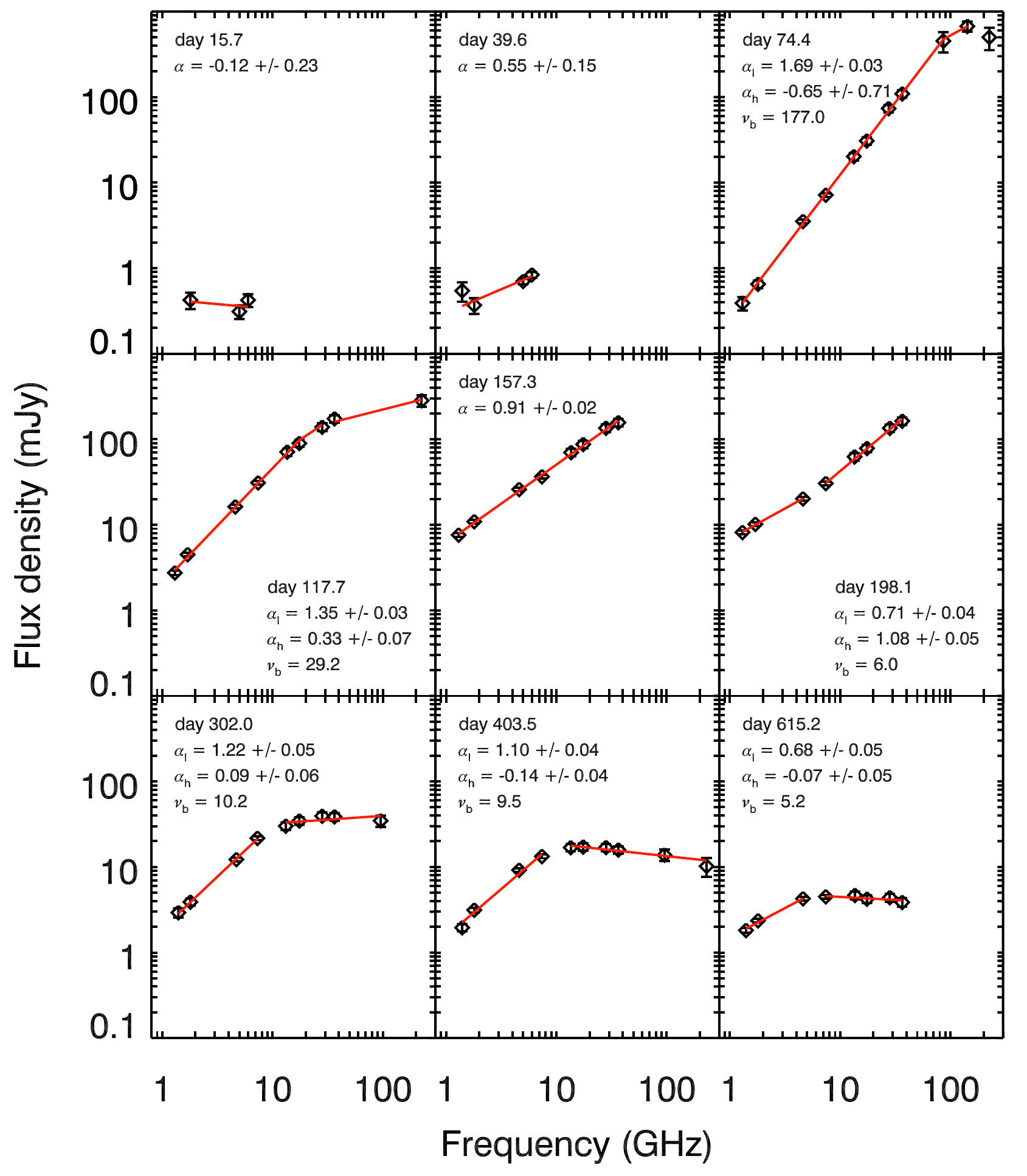}
\end{figure}

\begin{figure}
\centering
\includegraphics[width=6.2in, angle=0]{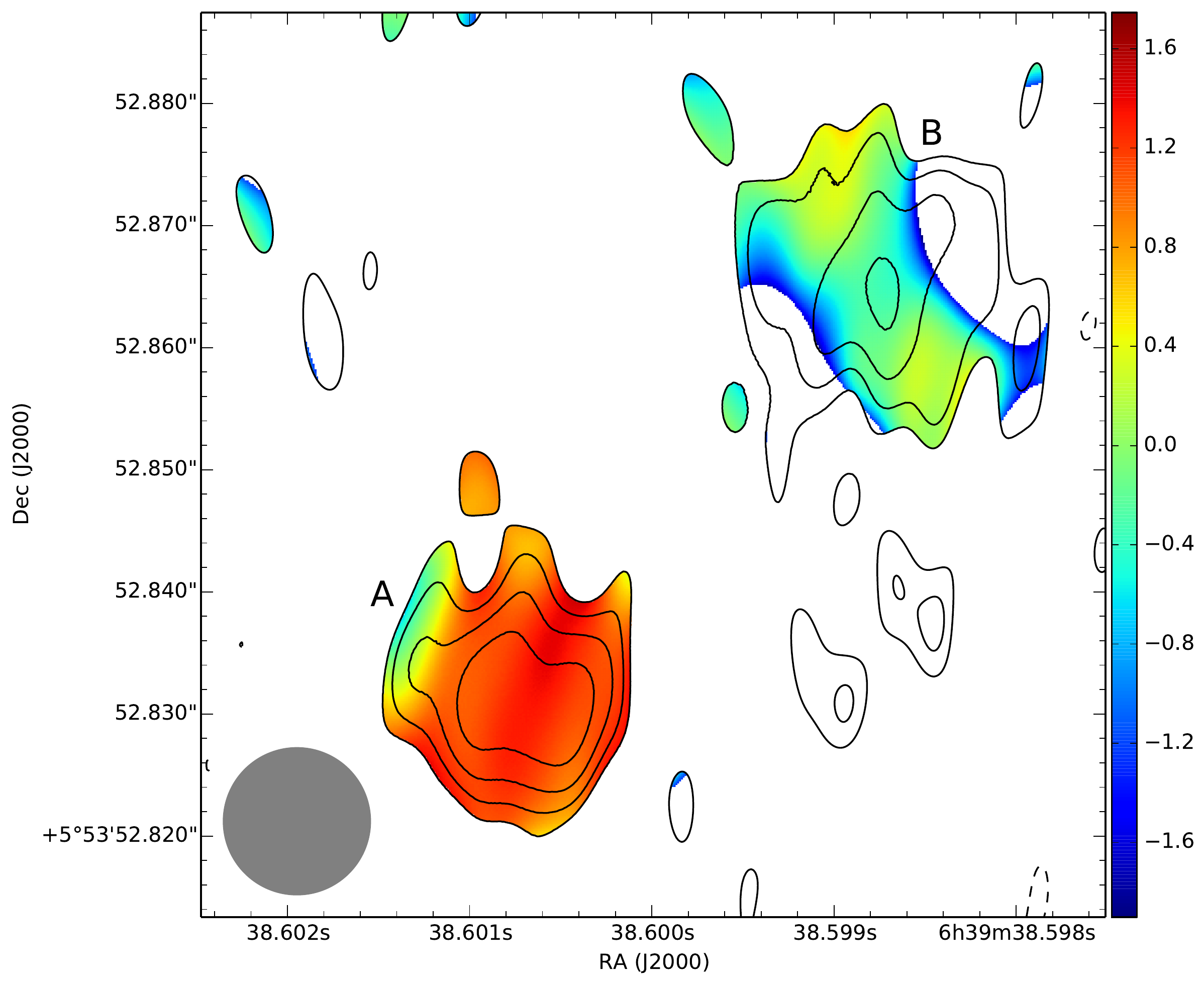}
\end{figure}

\begin{figure}
\centering
\includegraphics[width=6in, angle=0]{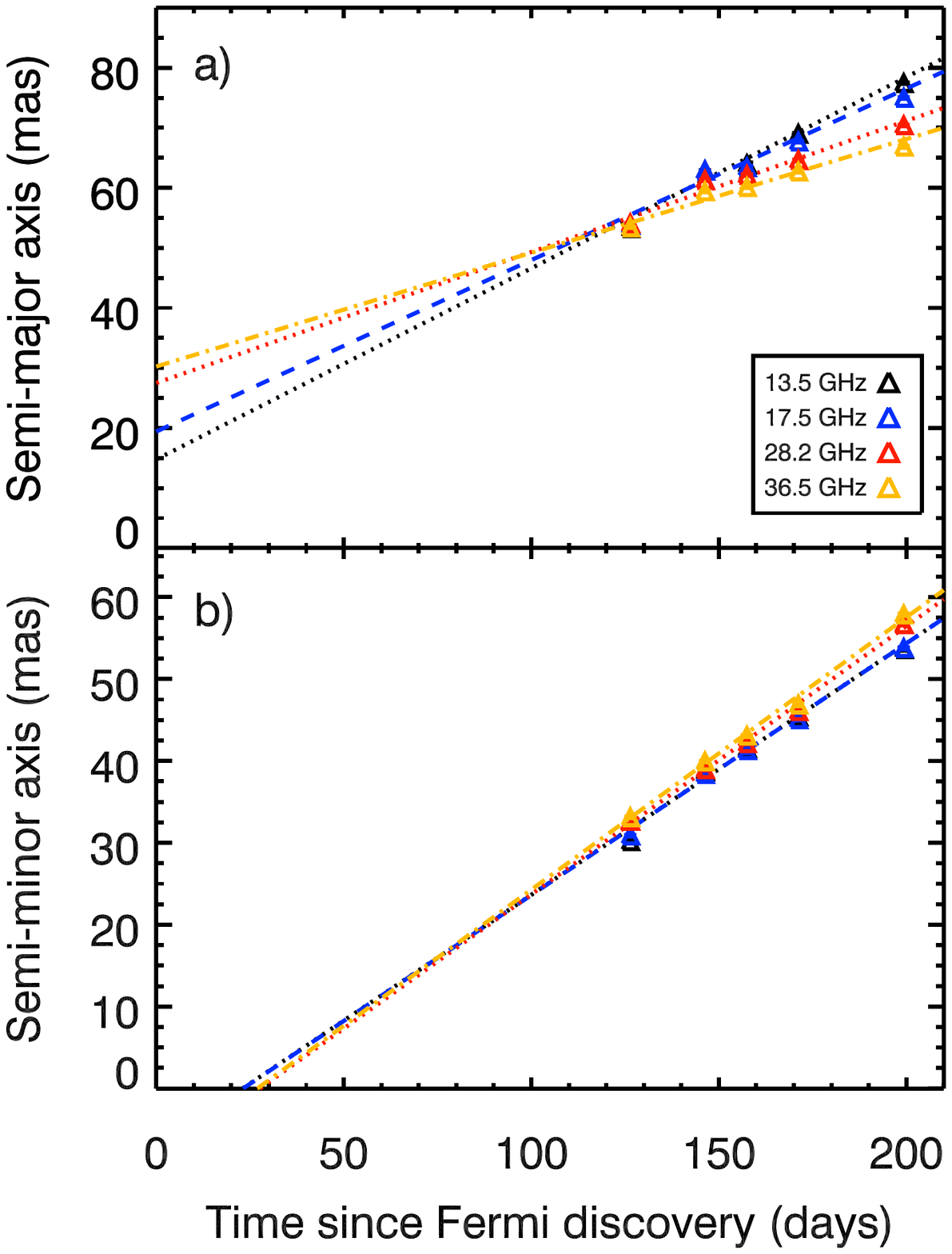}
\end{figure}

\begin{figure}
\centering
\includegraphics[width=6in, angle=0]{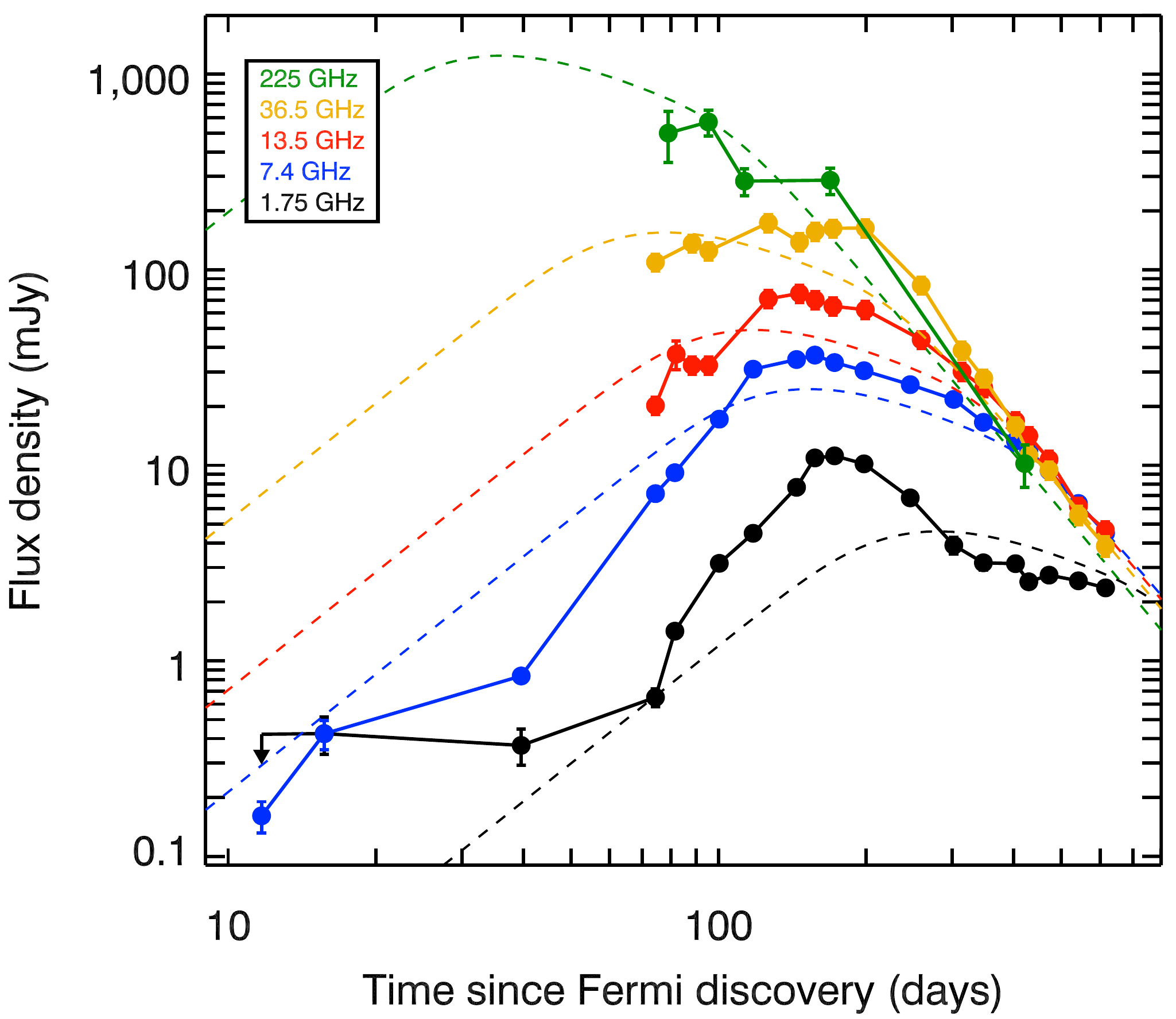}
\end{figure}



\clearpage
\setcounter{section}{0}
\pagestyle{plain}
\setcounter{page}{1}

\noindent {\bf {\Large Supplementary Information}}\\

\section{Radio/Millimeter Light Curve and Measurement of Ejecta Mass} 

Historically, the radio emission from novae has been understood as thermal emission from the warm, ionized, expanding ejecta\cite{Seaquist08, Roy13}. The radio emission starts out optically thick, with a spectral index $\alpha = 2$, and the flux rises as the photosphere expands. As the ejecta expand, the density drops, until the radio photosphere begins to recede back through the ejecta---first at high frequencies and later at low frequencies. This recession of the radio photosphere leads to the turn-over and eventual decline of the radio light curve. Eventually, the radio spectral index should match that of an optically-thin thermal plasma: $\alpha = -0.1$.

Selected frequencies of our VLA, SMA, and CARMA monitoring are plotted as the light curve in Fig.\ 1. We also show the radio spectral evolution in Fig.\ ED1. The spectrum was flat in the first two epochs before it steepened to what is expected for optically thick thermal emission. Later the radio spectrum turned over first at high frequencies, until at late times it had returned to flat.

The evolution of V959~Mon after Day $\sim$200 can be roughly fit with the simple picture of expanding thermal ejecta. In Fig.\ ED4, we show a simple model fit to the light curve as dashed lines using the ``Hubble Flow" model\cite{Hjellming79, Seaquist08}. This model assumes a homologous outflow with $v_{\rm max}$ = 2,400 km/s\cite{Ribeiro13}, a distance of 1.5 kpc\cite{Munari13}, and a $\rho \propto r^{-2}$ density profile. It also assumes that electron temperature is constant with both time and radius, a spherical geometry, and a filling factor of 0.1\cite{Shore13,Nelson14}. With these assumption, the light curve is best fit with an electron temperature $2\times10^4$ K, $v_{\rm min}$ = 200 km/s, and $M_{\rm ej} = 4 \times 10^{-5}$ M$_{\odot}$. The geometry of V959~Mon was clearly aspherical, and departures from spherical symmetry could decrease the ejected mass by as much as a factor of three\cite{Ribeiro14}. This ejecta mass estimate is well within the range of theoretical expectations and observed values for normal classical novae\cite{Gehrz98, Yaron05, Seaquist08}.

While much of the multi-frequency light curve of V959~Mon can be described with simple expanding thermal ejecta, there are a couple of notable departures from this model. The first is the unexpectedly bright, flat spectrum radio emission observed at early times (10--40 days after $\gamma$-ray discovery). These measurements cannot be explained with the free expansion of thermal nova ejecta, as will be discussed in more detail in the next section. The other striking departure from the thermal model is the evolution of V959~Mon at the lowest observing frequencies (1--2 GHz). As can be seen in Fig.\ ED4, the 1.75 GHz light curve shows a maximum around day 170, which is much brighter and more steeply peaked than expected for thermal emission. This low-frequency excess may be additional evidence for synchrotron emission in V959~Mon. The low-frequency behavior of classical novae remains poorly understood, as most novae have fallen below detection limits at lower frequencies. V959~Mon was an unusually nearby, radio-bright nova, enabling us to map its evolution at 1--2 GHz in detail; it remains to be established if its low frequency behavior is unusual amongst novae.

\section{Origin of the Early Flat-Spectrum Emission}
The flat radio spectrum seen during our first two epochs of radio observations (Days 12 and 16) was consistent with either synchrotron emission or optically thin thermal emission. 

\subsection{Synchrotron Emission}
While the standard optically-thin synchrotron spectrum is $\alpha = -0.7$, the spectrum of synchrotron emission observed in novae is often much flatter, $\alpha \approx -0.2$\cite{Mioduszewski04, Sokoloski08, Rupen08, Eyres09, Rupen14}, wholly consistent with our early measurements of V959~Mon.

We can estimate if $\sim$1--2 mJy of synchrotron emission is plausible in V959~Mon using the standard synchrotron formalism for supernovae\cite{Chevalier82, Chevalier98}. We assume that fractions of the postshock energy density are transferred to amplification of the magnetic field in the shock and acceleration of electrons to relativistic speeds ($\epsilon_B$ and $\epsilon_e$, respectively). The energy density in the magnetic field is therefore $U_B = \epsilon_B\, \rho\, v_s^2$, and the energy density in relativistic electrons is $U_e = \epsilon_e\, \rho\, v_s^2$. Here, $\rho$ is the density of the pre-shock material, and $v_s$ is the shock velocity. We use standard equipartition assumptions, $\epsilon_B = \epsilon_e = 0.1$. 

The observed flux densities can be produced via synchrotron radiation, even in the case of a relatively slow shock, $v_s$ = 2,000 km/s, as long as the pre-shock densities are high, $10^4-10^5$ cm$^{-3}$. Such high densities are expected in nova ejecta during the first months following outburst. Interaction between rapidly expanding polar ejecta and more slowly moving equatorial material could produce the differential velocities\cite{Livio90} and high density pre-shock material required for detectable synchrotron emission. 

\subsection{Optically Thin Thermal Emission} Here, we consider if optically thin thermal emission with flux density $\sim$0.4 mJy is plausible at such an early time in the outburst. The angular diameter of a thermally radio emitting body can be estimated as\cite{Seaquist08}: 
$$\theta = \sqrt{{1.36\, \lambda^2\, S_{\nu}\over{T_e\, \tau_{ff}}}}$$
where $\theta$ is in arcseconds, $\lambda$ is the observing wavelength in cm, $S_{\nu}$ is the measured flux density in mJy, $T_{e}$ is the source's electron temperature in K, and $\tau_{ff}$ is the free-free optical depth. We assume an electron temperature of 20,000 K, as derived from fitting the radio light curve and consistent with VLA A-configuration images from late 2012. If the source is to be described as thermal emission, it must be optically thin, so we estimate $\tau_{ff} \leq 0.1$. The most stringent constraint on the angular diameter comes from the 1.8 GHz detection on Day 16, from which we find $\theta \gtrsim 0.28^{\prime\prime}$ for the above assumptions.

At a distance of 1.5 kpc, this angular diameter corresponds to a physical diameter of $\gtrsim 6 \times 10^{15}$ cm. Assuming the outer ejecta started expanding on day 0 at 2,400 km s$^{-1}$, we would expect the diameter to not have exceeded $6.6 \times 10^{14}$ cm. Therefore, optically-thin thermal material of 20,000 K would have to be implausibly large to produce the observed flux density at 1.8 GHz. 

If the electron temperature was higher at earlier times, optically-thin thermally emitting material could have a smaller radius. 
Hot X-ray emitting material is observed in novae at a range of times following outburst, and is typically attributed to internal shocks in the ejecta\cite{Mukai08, Schwarz11}.
If we consider optically-thin thermal radio emission from shock-heated $5 \times 10^7$ K plasma, we find the diameter must have been $\gtrsim$6 mas, or $\gtrsim10^{14}$ cm. An optical depth of $\tau_{ff} = 0.1$ at $5 \times 10^7$ K and 1.8 GHz implies an emission measure of $10^{11}$ cm$^{-6}$ pc\cite{Seaquist08}. For a sphere of diameter $10^{14}$ cm, the implied number density filling the sphere was $5 \times 10^{7}$ cm$^{-3}$, with a total mass $5\times 10^{-8}$ M$_{\odot}$. The thermal energy in this material would have been $\sim 10^{42}$ erg.

These numbers are plausible, given the mass and energy budget of novae. Indeed, hard X-rays were detected from V959~Mon at the first opportunity, 61 days after $\gamma$-ray discovery\cite{Nelson12}. From these early time observations alone, we can not differentiate between synchrotron or shock-heated processes. In the next section, we compare measurements of the VLBI knots with concurrent X-ray observations to conclude that the shock-powered emission is, in fact, synchrotron radiation.

\section{Origin of the Compact VLBI Radio Knots}
The emission mechanism powering the compact VLBI knots can be additionally constrained by X-ray observations obtained around the time of our EVN and VLBA imaging. If the VLBI knots were due to free-free emission from hot shocked gas, they would also have emitted in the X-rays. V959~Mon was monitored with the X-ray Telescope (XRT) onboard the \emph{Swift} satellite throughout the months following outburst, and results have been published for two early epochs: 2012 Aug 19 (day 61)\cite{Nelson12} and 2012 Nov 18 (day 152)\cite{Nelson12a}.

On day 61, the X-ray spectrum was well fit by an absorbed thermal plasma\cite{Nelson12}, with absorbing hydrogen column density $N(H) = (3.1\pm0.4)\times10^{22}$ cm$^{-2}$, plasma temperature $T = \left(4.0^{+1.4}_{-0.8}\right)\times10^{7}$ K,
and observed (absorbed) 0.3--10 keV flux $F = 1.11 \times 10^{-11}$ erg s$^{-1}$ cm$^{-2}$. Assuming a distance of 1.5 kpc, the unabsorbed luminosity of this plasma was $1.0 \times 10^{34}$ erg s$^{-1}$.

On day 152, the X-ray spectrum showed evidence for a soft blackbody component (attributed to the photosphere of the hot white dwarf), in addition to the thermal plasma. The absorbed thermal plasma was best fit\cite{Nelson12a} with $N(H) = \left(8.5^{+0.2}_{-0.3}\right)\times10^{21}$ cm$^{-2}$, $T = (1.1 \pm 0.3) \times 10^{7}$ K, and $F = 5.4 \times 10^{-12}$ erg s$^{-1}$ cm$^{-2}$, implying an unabsorbed luminosity of $1.1 \times 10^{34}$ erg s$^{-1}$.

Assuming these X-rays were completely attributable to bremsstrahlung continuum, the volume emission measure ($EM_V = n^2\, V$, where $n$ is the number density of the gas and $V$ is the volume) of the hot plasma can be found by:
$$EM_V = 7.1\times10^{26}\ L\ T^{-1/2}$$
where $L$ is the luminosity in erg/s, $EM_V$ is the volume emission measure in units of cm$^{-3}$, and we assume a completely ionized hydrogen gas\cite{Rybicki79}. The emission measure was therefore $1.1 \times 10^{57}$ cm$^{-3}$  on day 61 and $2.4 \times 10^{57}$ cm$^{-3}$ on day 152. If lines contribute to the X-ray flux, these quantities will be upper limits on $EM_V$.

Radio optical depth scales as the path-length emission measure ($EM = n^2\, l$, where $l$ is the path length through the emitting gas, $l \approx V^{1/3}$), rather than $EM_V$. The two emission measures can be related as 
$$EM = 3.24 \times 10^{-19}\ EM_V\ V^{-2/3}\ f^{-2/3}$$
where $f$ is the volume filling factor of the gas and $EM$ is in units of cm$^{-6}$\, pc. The optical depth of radio emission from this gas can be expressed as\cite{Seaquist08}:
$$\tau_{\nu}= 2.67 \times 10^{-20}\ T^{-1.35}\ \nu^{-2.1}\ EM_V\ V^{-2/3}\ f^{-2/3}$$
where $\nu$ is the observing frequency in GHz. If we assume that the shock-powered radio and X-ray bremsstrahlung emission originate in three small knots, each of radius 5 mas or $10^{14}$ cm, we estimate an optical depth at 5 GHz of $\tau_{\nu} \approx 10^{-3}$ on day 61 and $\tau_{\nu} \approx 0.01$ on day 152. We note that such optically thin conditions were inconsistent with the spectral index measured for knot A on day 106 with VLBA ($\alpha \approx 1.2$).

As $\tau_{\nu} << 1$, the thermal radio flux density can be estimated as\cite{Seaquist08}:
$$S_{\nu} = 8.2 \times 10^{-4}\ T\ \theta^2\ \nu^{2}\ \tau_{\nu}$$
where $S_{\nu}$ is in units of mJy and $\theta$ is the angular size in units of arcsec. In this case, the dependence on $V$ and $f$ disappears, so that
$$S_{\nu} = 4.5 \times 10^{-18}\ T^{-0.35}\ \nu^{-0.1}\ EM_V\ d^{-2}$$
where $d$ is the distance in cm. We therefore find that the X-ray emitting plasma measured by \emph{Swift} would have yielded 5 GHz flux densities of 0.09 mJy on day 61 and 0.12 mJy on day 152. These values are an order of magnitude lower than the measured flux densities of the VLBI knots ($\sim$1.5--2 mJy; Table ED3) detected during days 91--117. This discrepancy implies that hot thermal plasma was not the dominant emission mechanism powering the VLBI knots. We therefore conclude that the VLBI knots emitted synchrotron radiation. By extension, we also assert that the flat-spectrum radio emission observed at early times was synchrotron radiation.

\end{document}